\documentclass[aps,pra,twocolumn,showpacs]{revtex4}  % for review and submission
\usepackage{graphicx}  % needed for figures
\usepackage{dcolumn}   % needed for some tables
\usepackage{bm}        % for math
\usepackage{amssymb,color}   % for math
\usepackage{textcomp}
\begin{document}

\title{Manifold approach for a many-body Wannier-Stark system: localization and chaos in energy space}

\author{C.~A.~Parra-Murillo and S. Wimberger}
\affiliation{Institut f\"ur Theoretische Physik and Center for Quantum Dynamics, Universit\"at Heidelberg, 69120 Heidelberg, Germany}

%\date{\today}

\begin{abstract}
We study the resonant tunneling effect in a many-body Wannier-Stark system, realized
by ultracold bosonic atoms in an optical lattice subjected to an external Stark force. The properties of 
the many-body system are effectively described in terms of upper-band excitation manifolds, which 
allow for the study of the transition between regular and quantum chaotic spectral statistics. We show that our 
system makes it possible to control the spectral statistics locally in energy space by the competition of the force
and the interparticle interaction. By a time-dependent sweep of the
Stark force the dynamics is reduced to a Landau-Zener problem in the single-particle setting.
\end{abstract}


\maketitle

\section{Introduction}\label{sec:1}

One of the most remarkable features of a quantum system is the ability of transporting electrons and 
atoms across classically forbidden regions. This process is allowed by means of a well known effect referred
to as quantum tunneling. In semiconductor physics \cite{Kittel}, the
use of a superlattice permits the enhancement of the transport 
along a specific spatial direction, supported, in a first approximation, by the resonant coupling between electronic levels.
Nowadays many of these solid state physics paradigms are amply investigated in a cleaner manner using ultracold atoms and optical potentials.
The rapid advances in the experimental techniques have opened a huge field of research, in which the most common approaches to the
description of many-particle physics are based on Bose-Hubbard-type
Hamiltonians \cite{Zoller1998,IblochREP2008,IblochREP2008-1}. A basic
feature of these Hamiltonian models is that the kinetic energy is
described by means of dynamical hoppings (tunneling) between different potential wells.
Here, we are extending such a model to higher energy bands coupled by means of an external Stark force.

The study of many-body effects at resonant tunneling conditions is not straightforward since, in the presence of strong correlations, the complexity
is overwhemingly increased
\cite{kol68PRE2003,TomadimPRL2007,TomadimPRL2008}. For the single-particle and mean field limits, 
there are up to date experimental realizations of the Wannier-Stark
system (see for instance \cite{WimbPRL2007,WimbPRL2009,Oliver,Oliver-1,Oliver-2,Tayebirad2011,NilsPRA,ChSalomonPRL1997,Morsch2001,ChNaegerl,Raizen1996}). Additionally, 
by investigating certain parameter regimes, for example, in the case of strongly interacting atoms (hard core bosons), an effective   
description is possible by mapping the many-body lattice system to
analytically solvable effective Hamiltonians \cite{PloetzEPJD2011,MGreiner2011-0,InsbruckArxiv2013}.
In this paper, we investigate the resonant tunneling effect in a Bose-Hubbard model extended to a second excited Bloch band. This
system, that can be immediately realized in experiments (see ref.~\cite{CarlosPRA2013,CarlosThesis2013} for details), has very
 interesting spectral properties. These allow for the study of many-body effects as, for example, interaction-induced 
quantum chaos \cite{kol68PRE2003,TomadimPRL2007}, diffusion and relaxation in Hilbert space \cite{CarlosPRA2013} and coherent dynamics
in the weak interacting regime \cite{PloetzEPJD2011,PloetzJPB2010}. We show that the main spectral properties can be captured in an effective
theory based on upper-band manifold excitations, that can, in principle, be measured in experimental realizations such as reported in \cite{,MGreiner2011-1}.
Our approach allows us to characterize: the onset of quantum chaos, to distiguish the condition for the emergence of localization in
energy space \cite{IzrailevPhysSkripta2001,APolkovnikov2013}, and to design the type of driving dynamics that can be implemented in 
analogy with the well-known Landau-Zener process
\cite{KorschPRep2002,WimbPRL2007,WimbPRL2009,Oliver-1,NilsPRA,LZ,Arimondo2011,Tayebirad2011}. 

\section{The Many-Body Wannier-Stark System}\label{sec:2}
\subsection{The System and the Two-band Model}\label{sec:2a}

Our system consists of ultracold bosonic atoms in a optical lattice \cite{IblochREP2008,IblochREP2008-1} 
subjected to an external Stark force. The force stimulates $(a)$ the atomic transport along the lattice, for instance, atomic Bloch oscillations 
\cite{ChSalomonPRL1997,Morsch2001,ChNaegerl}, and $(b)$ between the Bloch bands, e.g. Landau-Zener transitions \cite{Oliver,Oliver-1,Oliver-2,NilsPRA,Tayebirad2011}. This latter 
process is characterized by the exchange of particles between the bands and is enhanced at specific values of force, $F_r\approx \Delta_g/2\pi r$, 
where $\Delta_g$ is the energetic gap between the two bands. At those values, resonantly enhanced tunneling takes place between Wannier-Stark
levels distancing $r$ wells \cite{WimbPRL2007,PloetzJPB2010,PloetzEPJD2011,CarlosThesis2013}. The integer $r$ is from now on called the order
of the resonance. In the following, we restrict to two coupled energy bands. Such a situation is realized, e.g., in a double-periodic lattice,
see ref.~\cite{CarlosPRA2013}.

The many-body physics can then be described by the celebrated Bose-Hubbard model extended to a two-band scenario 
\cite{CarlosThesis2013,CarlosPRA2013}. The corresponding Hamiltonian reads
\begin{eqnarray}\label{eq:01}	
 \hat{H} &=&\sum^L_{l=1}\sum_{\beta}\left[-\frac{J_{\beta}}{2}\left(\hat{\beta}^{\dagger}_{l+1}\hat{\beta}_{l}+h.c.\right)+
 \frac{W_{\beta}}{2}\hat{\beta}^{\dagger 2}_{l}\hat{\beta}^2_{l}\right]\,\nonumber\\
 &+&\sum^L_{l=1}\sum_{\mu} \omega_B C_{\mu} (\hat{a}^{\dagger}_{l+\mu}\hat{b}_l+h.c.)+\sum_{\beta}\varepsilon^{\beta}_l\hat{n}^{\beta}_l\,\nonumber\\
 &+&\sum^L_{l=1}2W_x\hat{n}^a_l\hat{n}^b_l+
 \frac{W_x}{2}\left(\hat{b}^{\dagger 2}_{l}\hat{a}^2_{l}+h.c.\right)\,.
\end{eqnarray}

The annihilation (creation) operators are defined by ${\beta}_l({\beta}^{\dagger}_l)$, and the number operators 
are $n^{\beta}_l={\beta}^{\dagger}_l{\beta}_l$, with band index defined as $\beta=\{a,b\}$. The hopping amplitudes are $J_{\beta=a,b}$, and the on-site 
interparticle interaction per band has a strength $W_{\beta}$. The interband coupling is generated by dipole-like couplings
($C_{\mu}$) and by the repulsive interaction ($W_x$). The large dimensionality of the parameter space makes it impossible to analitically solve 
our problem. Therefore a numerical procedure to find the eigensystem is required. We quickly summarize the
procedure in the following, and refer the reader to ref.~\cite{CarlosThesis2013} for details.  

\subsection{Numerical Treatment}\label{sec:2b}
In order to study the eigenenergy spectrum of the Hamiltonian (\ref{eq:01}), we first implement a transformation into the interaction picture with respect
to the external force. In this procedure the term $\sum_{l,\beta}\omega_Bl\hat{n}^{\beta}_l$ is removed, and the hopping and dipole-like terms transform as:
$\hat{\beta}^{\dagger}_{l+1}\hat{\beta}_l\rightarrow\hat{\beta}^{\dagger}_{l+1}\hat{\beta}_l\exp{(-i\omega_Bt)}$ and 
$\hat{a}^{\dagger}_{l+\mu}\hat{b}_l\rightarrow\hat{a}^{\dagger}_{l+\mu}\hat{b}_l\exp{(-i\omega_B \mu Ft)}$. The big advantage of the transformation into 
the interaction picture is that now the new Hamiltonian is invariant under translation and periodic in time. Its period is given by $T_B=2\pi/\omega_B$, 
which is known as the Bloch period. We now set periodic boundary conditions in space, i.e. $\hat{\beta}^{\dagger}_{L+1}=\hat{\beta}^{\dagger}_{1}$. In order
to diagonalize the time-dependent Hamiltonian, $\hat H(t)$, we use the translationally invariant Fock states $\{|\gamma_i\rangle\}$ defined in 
Refs.~\cite{kol68PRE2003,TomadimPRL2007}, with dimension given by 
\begin{equation}\label{eq:02}
\mathcal{N}_s=\frac{(N+2L-1)!}{LN!(2L-1)!},
\end{equation}
\noindent where $L$ is the number of lattice sites and $N$ the total
particle number. We study the eigensystem of the Floquet Hamiltonian \cite{Shirley1965} $\hat{H}_f=\hat{H}(t)-i\partial_t$, whose eigenenergies $\varepsilon_i$ are defined as the 
set of eigenvalues of $\hat H_f$ lying within the Floquet zone $\varepsilon_i\in [-\omega_B/2,\omega_B/2]$ (see details of the diagonalization procedure
in \cite{CarlosThesis2013}). In the following we introduce an effective, analytical description of the spectral properties based on
the upper-band excitation and its respective comparison with the exact numerical results.

\subsection{Manifold Approach}\label{sec:2c}

The non-interacting limit is described by the Hamiltonian Eq.~(\ref{eq:01}) with $W_{a,b,x}=0$. Here it is possible to construct a local Hamiltonian around a
single resonance $r$. This implies that the site $l_a$ may be connected to the
upper-band lattice site $l_b=l_a-r$ ($r$ sites to the left). We now rescale the Hamiltonian as $\hat{H}\rightarrow\hat{H}/\Delta_g$, where
$\Delta_g$ is typically the largest parameter; thus we have $|C_{|\mu|>0}|/\Delta_g\ll 1$, which means that only working with the largest 
dipole strength, $C_0$, is enough to capture the essential features of
the system. The effective Hamiltonian around the resonance of order $r$ reads
\begin{eqnarray}\label{eq:03}
\hat{H}_r&=&\sum^L_{l=1}\sum_{\beta}\left[-\frac{J_{\beta}}{2}\left(\hat{\beta}^{\dagger}_{l+1}\hat{\beta}_{l}+h.c.\right)
	+(\Delta_{\beta}-\omega_B r)\hat{n}^{\beta}_l\right]\nonumber\\
&+&\sum^L_{l=1}\omega_BC_0\left(\hat{a}^{\dagger}_{l}\hat{b}_l+h.c.\right),
\end{eqnarray}

\noindent where the energy separation between the Bloch bands is $\Delta_{a,b}=\{0,\Delta_g\}$. We then have a Hamiltonian consisting of two
tilted lattices. The eigenenergies of the independent lattices are given
by the Wannier-Stark ladder formula \cite{KorschPRep2002}:
\begin{equation}\label{eq:04}
 \varepsilon^{\beta}_{l}=\omega_B l_{\beta} +\Delta_{\beta},\;\;\;\text{with} \;\;\;l_{\beta}\in\mathbb{Z}\,.
\end{equation}

In the case $F=0$, that is, without inter-band coupling and no tilt, the eigenstates of Eq.~(\ref{eq:04}) can be classified according
to their number of particles in the upper Bloch band, defined as: 
\begin{equation}
M_i=\langle\varepsilon_i|\hat M|\varepsilon_i\rangle,\;\;\;\text{with}\;\;\; \hat M=\sum\nolimits_l\hat n^b_l.
\end{equation}
 By writting the Hamiltonian matrix representation (\ref{eq:04}) in the basis $|\gamma_i\rangle$, ordered by increasing upper-band occupation number $M$, the 
 Hamiltonian is reduced to the block matrix
\[
\hat H_r=\left(
\begin{array}{cccc}
 \mathbf{H}_{0,0} & \mathbf{H}^{\dagger}_{0,1} &                    &                              \\
 \mathbf{H}_{0,1} & \mathbf{H}_{1,1}           & \ddots             &                              \\
                  & \ddots                     & \ddots             & \mathbf{H}^{\dagger}_{0,M-1} \\ 
                  &                            & \mathbf{H}_{0,M-1} & \mathbf{H}_{M,M}             \\ 
\end{array}
\right)
\label{eq:05}.
\]

\begin{figure}[t]
\centering \includegraphics[width=\columnwidth]{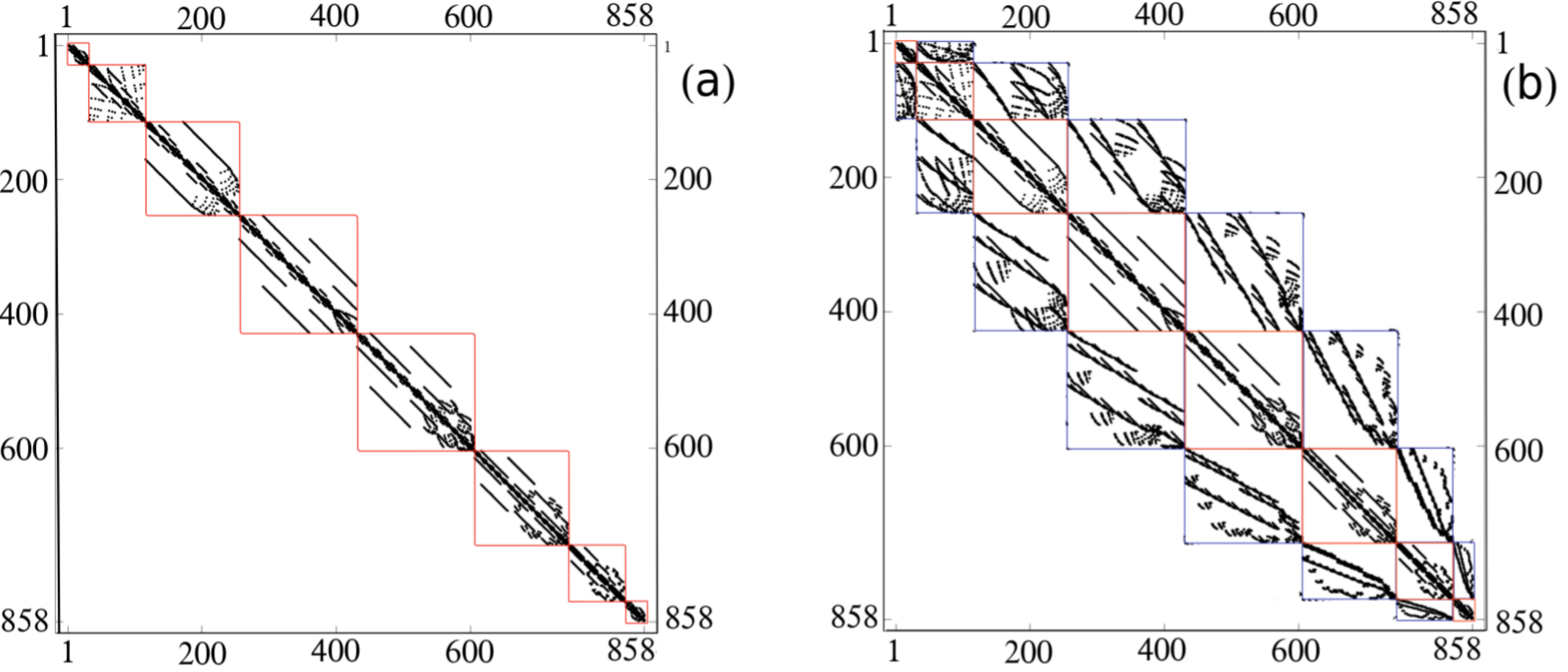}
\caption{\label{fig:1}(Color online): Hamiltonian matrix written in the $|\gamma_i\rangle$ basis. (a) 
Block structure of the Hamiltonian in the non-interacting case, for
the system $N/L=7/4$, with parameter $F_{r=1}=0.25$.
 Since $N=7$ the matrix contains $8$ blocks corresponding to the $N+1$ manifolds. The intra-manifold off diagonal couplings
are set by the hopping terms that do not couple $\gamma_i$ states with different manifold number. (b) Tight-binding-type many-body
Hamiltonian matrix for the same parameter from panel (a). The other parameters are: $\Delta_g=1.61$, $J_a=0.082$, $J_b=-0.13$,
$C_0=-0.094$, $C_{\pm 1}=0.037$, and $C_{\pm 2}=-0.0022$.}
\end{figure}

\noindent The ${\rm diag}(\hat{H}_r)=\oplus^N_{M=0}(\hat H_r)_{M}$ is a diagonal block matrix constructed through the Hamiltonian terms preserving 
number $M$, i.e., the hopping and energy terms in Eq.~(\ref{eq:01}). The blocks on the diagonal are matrices with
dimension $d_M\times d_M$ (see Fig.~\ref{fig:1}(a)), where $d_M$ is given by 
\begin{eqnarray}\label{eq:06}
d_M=\frac{1}{L}{M+L-1\choose L-1}{N-M+L-1\choose L-1}.
\end{eqnarray}

\noindent The Hamiltonian only contains non-zero coupling between those inter-manifold states with excess of particles $\Delta M=\pm 1$,
with "hopping" strength $\omega_BC_0$ (see Fig.~\ref{fig:1}(b)). In addition, the Hamiltonian can be reduced to a tight-binding-type one for
the upper-band excitation manifolds (from now on labeled by $M$) when
describing the averaged one-particle exchange processes in the
resonant system. To do this, we use the closure relation $\sum_{i}|\gamma_i\rangle\langle\gamma_i|=\mathbf{1}$, which can be rewritten in terms of the manifold projectors $\hat P_M$ as follows 
\begin{equation}\label{eq:07}
 \sum^{N}_{M=0}\hat{P}_M=\hat\mathbf{1}\,,\;\;\;\;\text{with}\;\;\;\;\hat{P}_M=\sum^{d_M}_{i=1}|\gamma_i;M\rangle\langle\gamma_i;M|.
\end{equation}

\noindent Here
$|\gamma_i;M\rangle\equiv|n^a_1,n^a_2,...\rangle\otimes|n^b_1,n^b_2,...\rangle$
is a Fock state with $M$ particles in the upper band. The 
projectors $\hat{P}_M$ allow one to rewrite the Hamiltonian as $\hat\mathbf{1}\hat H_r \hat\mathbf{1}=\sum_{M,M'} \hat P_M\hat H_r\hat P_{M'}$
with the eigenstate $|\psi\rangle$ of $\hat H_r$ expanded as
\begin{eqnarray}\label{eq:08}
|\psi_M\rangle=\hat{P}_M|\psi\rangle=\sum_i|\gamma_i;M\rangle\langle \gamma_i;M|\psi\rangle,
\end{eqnarray}

%******** FIG 2

%******** FIG 3
\noindent and $\langle\psi_{M'}|\psi_M\rangle=\delta_{M,M'}$. The off-diagonal blocks are not square matrices but their dimension is 
$d_{M}\times d_{M+1}$. They are computed from the single-particle exchange term 
\begin{equation}\label{eq:09}
\hat{H}_{M,M'}\equiv\langle\psi_{M'}|\sum^L_{l=1}\omega_BC_0\left(\hat{a}^{\dagger}_{l}\hat{b}_l+h.c.\right)|\psi_M\rangle.
\label{eq:fmanifold}
\end{equation}

By choosing $\langle\gamma_i;M|\psi\rangle=1/\sqrt{d_M}$ we obtain a
simplified tight-binding-type Hamiltonian for the manifolds
\begin{eqnarray}\label{eq:10}
 \hat{H}'_{r}\simeq\sum^{N}_{M=0}\varepsilon^r_M |\psi_{M}\rangle\langle \psi_{M}|+
 \omega_BC_0(|\psi_{M}\rangle\langle \psi_{M+1}|+h.c.),\;\;\;
\end{eqnarray}

\noindent where $\varepsilon^r_M=(\Delta_g-\omega_B r)M+(J_a-J_b)M$, and $\tilde\omega_B\equiv\omega_B C_0\sqrt{M+1}$.
Here we used the relation $N=N_a+N_b$, with $M\equiv N_b$, and the order of the resonance is approximately given by $r\approx\Delta_g/\omega_B$. 
To obtain the Hamiltonian (\ref{eq:10}) we have assumed that there is no additional relevant subclass of Fock states in any $M$-subspace,
and all possible single-particle processes are equally probable. Under this condition, the Hamiltonian (\ref{eq:10}) averages over  
hopping and dipole-like transition processes, and its final dimension is just $N+1$. 
A different choice of the distribution of the coefficients $\langle\gamma_i;M|\psi\rangle$ would lead to a similar effective Hamiltonian, 
but restricted to fewer participating states.

Interestingly, from the new Hamiltonian we can easily 
recognize an emerging localization of its respective eigenfunctions $|\phi\rangle$. To see this, we use 
$|\phi\rangle=\sum\nolimits_M Q_M|\psi_M\rangle$, which together with the Schr\"odinger equation, $\hat H'_r|\phi\rangle=E|\phi\rangle$, 
yields the coefficient equation: 
\begin{equation}\label{eq:11}
(E-\epsilon M)Q_M=\tilde{\omega}_B(Q_{M+1}+Q_{M-1}),
\end{equation}

\noindent with $\;\;\epsilon=\Delta_g-\omega_B r+J_a-J_b$. This equation can be solved using the ansatz $Q_M=AJ_{M'-M}(x_B)$, where 
$x_B\equiv 2\omega_BC_0/\epsilon$ and $J_{M'-M}(x_B)$ is the Bessel function of the first kind. Therefore, by using the 
identity $2k J_k(x)=x(J_{k+1}(x)+J_{k-1}(x))$ we find that the solution of Eq.~(\ref{eq:11}) is: $E_M=\epsilon M$ 
and $|\phi_M\rangle=A\sum\nolimits_{M'}J_{M'-M}(x_B)|\psi_{M'}\rangle$, with $A$ being a normalization constant. In energy space the
eigenfunction can be written as  
\begin{equation}\label{eq:12}
\phi_M(\varepsilon)\equiv\langle\varepsilon|\phi_M\rangle=A\sum\nolimits_MJ_{M'-M}(x_B)\psi_{M'}(\varepsilon)
\end{equation}
\noindent where $\psi_{M'}(\varepsilon)$, the eigenfunctions of the Hamiltonian $\hat H'_r(\tilde{\omega}_B=0)$, is a well-localized
function around the energy $\varepsilon^r_M$. These functions clearly satisfy the relation $\psi_{M}(\varepsilon-\varepsilon^r_{M'})=\psi_{M'}(\varepsilon)$.
These are Wannier-like functions in energy space. The probability density is then now given by 
\begin{equation}\label{eq:13}
\frac{|\phi_M(\varepsilon)|^2}{|A|^2}=|J_0(x_B)\psi_M(\varepsilon)+\sum_{M'\neq M}J_{M'-M}(x_B)\psi_M(\varepsilon)|^2,
\end{equation}
\noindent which means that, for $x_B\ll 1$, the probability maximizes around the manifold energy $\varepsilon^r_M$. The condition for this
to happen is to be far from the resonance where for typical system parameters: $2\omega_BC_0\ll \Delta_g-\omega_Br+J_a-J_b$ . At the resonance
we have $\Delta_g\approx\omega_Br$ and $x_B\approx 1$, which implies an overlapping of neighbor manifolds, since the expansion coefficients
in Eq.~(\ref{eq:13}) with $|M'-M|>0$ become non-negligible. This introduces a kind of hybridization effect between the manifolds
responsible for the destruction of the strong localization of the eigenfunctions $\phi_M(\varepsilon)$
(see refs.~\cite{IzrailevPhysSkripta2001,APolkovnikov2013} for other contexts of localization in energy space). 

The energy gap between two neighboring manifolds characterizes the one-particle exchange process (see Fig.~\ref{fig:3}(b)) and can be estimated
by straightforward diagonalization of the two-level Hamiltonian matrix
\begin{eqnarray}\label{eq:14}
H^r_{2\times 2} =
\left(
\begin{array}{cc}
\varepsilon^r_{M+1} & \omega_BC_0 \\
\omega_BC_0        & \varepsilon^r_M  \\
\end{array}
\right) \,,
\end{eqnarray}
\noindent from which we obtain 
\begin{eqnarray}\label{eq:15}
 \Delta_r=\Delta_g\sqrt{\left(1-\frac{\omega_B r}{\Delta_g}+\frac{J_a-J_b}{\Delta_g}\right)^2+4\left(\frac{\omega_B C_0}{\Delta_g}\right)^2}.\;\;\;\;\;
\end{eqnarray}

\noindent We notice that the minimal energy range of the many-level spectrum is thus given by $\Delta E=N\Delta^{\rm min}_r$, with 
$\Delta^{\rm min}_r\approx2\omega^r_B|C_0|$ and $\omega^r_B\equiv2\pi F_r$  (for typical parameters, $\Delta_g,\omega_B\gg |J_b-J_a|$). This
energy scale is shown in Fig.~\ref{fig:2}(a) for the single-particle case ($N/L=1/10$). 

\subsection{Effects of the Interparticle Interaction}\label{sec:2d}
\begin{figure}[t]
\centering \includegraphics[width=0.8\columnwidth]{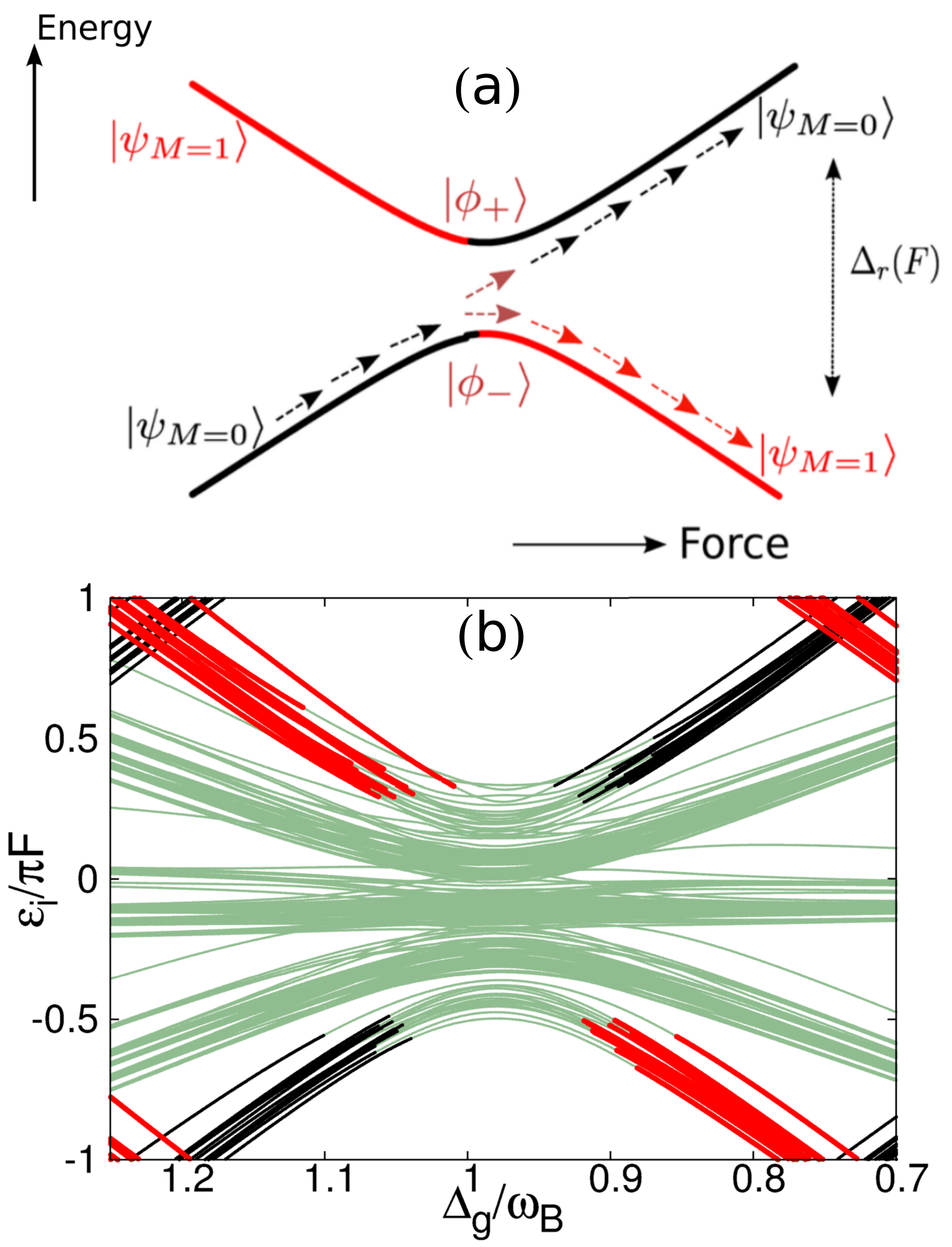}
\caption{\label{fig:2}(Color online): (a) Numerically computed energy
  spectrum versus the Stark force in the single-particle case around
  $r=1$, with $N/L=1/10$. The inter-band coupling manifests itself around the resonance position $F_r$ by an avoided crossing.
(b) Many-body spectrum for $N/L=4/5$, around $r=1$. The different
color lines correspond to eigenstates of the type:
$|\psi_{M=0}\rangle$ (black), $|\psi_{M=1}\rangle$ (red), and the mixed-like 
states, i.e., states with $0<M<N$ are represented by the green
lines. In panel (a) $|\phi_{\pm}\rangle$ represent
the hybridized states at $F=F_r$. The parameters are those of
Fig.~\ref{fig:1} with additional interaction strengths: $W_a=0.021$,
$W_b=0.026$ and $W_a=0.023$.}
\end{figure}

A more precise description of the many-body spectrum is obtained when considering the effects of the interactions, i.e., for $W_{a,b,x}\neq0$. 
This induces a splitting of the internal manifold levels and couplings
between the $|\gamma_i\rangle$ states. The coupling is strong especially at
resonant tunneling condition (e.g. $F\approx F_r$). In this region, the manifold levels come closest (see Fig.~\ref{fig:2}(b)) and a natural
mixing of the manifold states occurs, that is, the eigenstates of (\ref{eq:01}) become hybridized as explained in Sec.~\ref{sec:2c}. This
latter effect is associated with the occurrence of avoided crossings
(ACs) around $F_r$, i.e., with the lack of symmetries in the system \cite{CarlosThesis2013}. We can
now easily estimate the largest manifold splittings generated by interparticle interaction. This is done by considering the basis states
$\{|\gamma_i\rangle\}$ with $M$ particles sitting in a single-particle level, in one lattice site, for example $|\gamma_i\rangle\sim|N-M,0,..\rangle_a\otimes|M,0,...\rangle_b$. The energy cost due to the interaction strengths $W_{a,b,x}$ is thus given by
\begin{eqnarray}\label{eq:16}
  (U^M_a)_{\rm max}&=&\frac{W_a}{2}(N-M)(N-M-1),\nonumber\\
  (U^M_b)_{\rm max}&=&\frac{W_b}{2}M(M-1),\nonumber\\
  (U^M_{ab})_{\rm max}&=&2W_x(N-M)M,
\end{eqnarray}

\noindent which allow one to compute the maximal intra-manifold
splitting as $U(M)\equiv\max\{(U^M_{\beta=a,b})_{\rm
  max},(U^M_{ab})_{\rm max}\}$. We can also rewrite the width $\Delta E$ as follows
\begin{eqnarray}\label{eq:17}
\Delta E=N\Delta^{\rm min}_r+U(N).
\end{eqnarray}
This expression is obtained by considering the maximal splitting of the highest manifold $M=N$, which occurs for those eigenstates, whose maximal projection in the $|\gamma_i\rangle$ space is given by the state with $N$ particles in the same upper-band level, say, the state
$|\gamma_i\rangle\sim|000,...\rangle_a\otimes|N00,...\rangle_b$.

%******** FIG 2
%******** FIG 3

In general, far from the resonance, any eigenstate of Eq.~(\ref{eq:01}) can
be rewritten in the dressed-like state basis $\{|M,\theta_{a},\theta_{b},\theta_{x}\rangle\}$ characterized by the integers
$\theta_{\beta}=\langle\varepsilon_i|\sum\nolimits_l\hat n^{\beta}_l(\hat
n^{\beta}_l-1)/2|\varepsilon_i\rangle$  and
$\theta_{x}=2\langle\varepsilon_i|\sum\nolimits_l\hat n^{a}_l\hat
n^{b}_l|\varepsilon_i\rangle$. The eigenenergies
 can then be approximated by 
\begin{equation}\label{eq:18}
\varepsilon_i(M,\vec{\theta})\approx M_i\Delta_r+W_a\theta_{a,i}+W_b\theta_{b,i}+W_x\theta_{x,i}. 
\end{equation}

The mixing between the different $M$-manifolds in the presence of inter-particle interaction is now also triggered by the coupling between inter-manifold
states with excess $\Delta M=\pm 2$. This fact implies that the Hamiltonian Eq.~(\ref{eq:10}) contains a new term, which is just a second
neighbor transition from the manifold $M$ to the manifold $M\pm2$. This is nothing but an extended tight-binding model,
with increasing on-site energies $\varepsilon_i(M,\vec\theta)$. Therefore, even in the presence of weak
interactions, the eigenstates $\langle\varepsilon|\phi_M\rangle$ preserves localization features. 

Computing the following commutator
\begin{eqnarray}\label{eq:19}
\left[\frac{\hat{H}}{\Delta_g},\hat{M}\right]=\sum\nolimits_{l,\mu}
\frac{\omega_B C_{\mu}}{\Delta_g}\hat{a}^{\dagger}_{l+\mu}\hat{b}_{l}
+\frac{W_x}{2\Delta_g}\hat{a}^{\dagger2}_{l}\hat{b}^2_{l}-h.c.\;\;
\end{eqnarray}

\noindent shows that the transition from weak to strong mixing is determined by the competition between one- and
two-particle exchange between the bands. This competition is the stronger the closer we are in resonance, at which we have that
$\omega_B/\Delta_g\approx 1/r$. Furthermore, it is expected that for a filling factor 
$N/L\sim 1$, single- and two-particle transitions have the same occurrence probability, which makes the system strongly
mixed. In the case $N/L\ll 1$, the dominant effect is a one-particle exchange characterized by the energy scale $\Delta_r$, and similarly for $N/L\gg1$,
the two-particle exchange process dominates with energy scale proportional to $W_x$. These two latter cases favor the weak manifold mixing.
Therefore, the eigenstates of (\ref{eq:01}) are expected to be localized in energy space according to the effective Hamiltonian (\ref{eq:10}).

So far, we have studied the properties of the Hamiltonian
(\ref{eq:01}) by using rather simple approximations based on the concept of manifold
excitation. The results have shown to be in a good agreement with the numerical ones. We now show two direct consequences
of the spectral properties; first we briefly describe the implementation of Landau-Zener-type dynamics by making the Stark force time-dependent.
In this way, driving individual eigenstates $|\varepsilon_i\rangle$ across an avoided crossing may be straightforwardly implemented. Secondly, we study the 
connection between strong manifold mixing and quantum chaos in the resonant tunneling regime.

\section{Numerics: Landau-Zener Dynamics}\label{sec:3}

Around a local resonance, our system provides a natural scenario for the study of Landau-Zener-type transitions. 
This can be done by defining a pulse $F(t)=F_0 + \alpha t$, with $\alpha=\Delta F/\Delta T$. Here $\Delta F$ is the effective region of
resonant tunneling and $\Delta T$ is the sweeping time. We focus on the dynamical driving of a state from the lowest manifold
$|\psi_{M=0}\rangle$ for the single-particle case. We have then two manifolds $M=\{0,1\}$ and the spectrum of this system is shown in 
Fig.~\ref{fig:1}(a). We define the sweeping rate $\alpha$ using the Heisenberg relation $\Delta Td \approx 1$, where 
$d=\Delta E/\mathcal{N}_s=\Delta_r$ is mean level spacing at $F_r$. 

The Hamiltonian is now time-dependent, $H(F(t))$, and the temporal evolution is computed by using a fourth-order Runge-Kutta method. 
In analogy with the LZ problem, the evolution across an avoided crossing can be diabatic, non-adiabatic and adiabatic. In the current case the 
dynamical regimes are determinated by the parameter $\lambda\equiv\alpha/d\,\Delta F$. Thus for $\lambda\ll1$ we have slow driving
through resonances, i.e., an adiabatic passage; for $\lambda\gg1$ a diabatic one, also referred to as sudden quench, and for $\lambda\sim 1$ we have
a non-adiabatic evolution. 

To follow the evolution of a state across the RET regime, we compute the detection probabilities 
$p_i=|\langle\varepsilon_i|\hat U(t)|\psi(0)\rangle|^2$, where $\hat U(t)=\hat{\mathcal{T}}\exp\left[-i\int_0^{t} \hat{H}(F(t))dt\right]$.
The distribution probability of the evolved wavefunction in the local energy space can be represented by means of the 
local density of states \cite{TDittrich1991,LFSantosPRL2012}:
\begin{equation}\label{eq:20}
P_{\psi}(\varepsilon,t)=\sum\nolimits_{i}p_i\delta(\varepsilon-\varepsilon_i),
\end{equation}

\noindent with the delta function defined as
\begin{equation}\label{eq:21}
 \delta(\varepsilon-\varepsilon_i)=\lim_{\delta\rightarrow 0}\frac{1}{\pi}\frac{\delta}{(\varepsilon-\varepsilon_i)^2+\delta^2}.
\end{equation}

\begin{figure}[t]
\centering \includegraphics[width=\columnwidth]{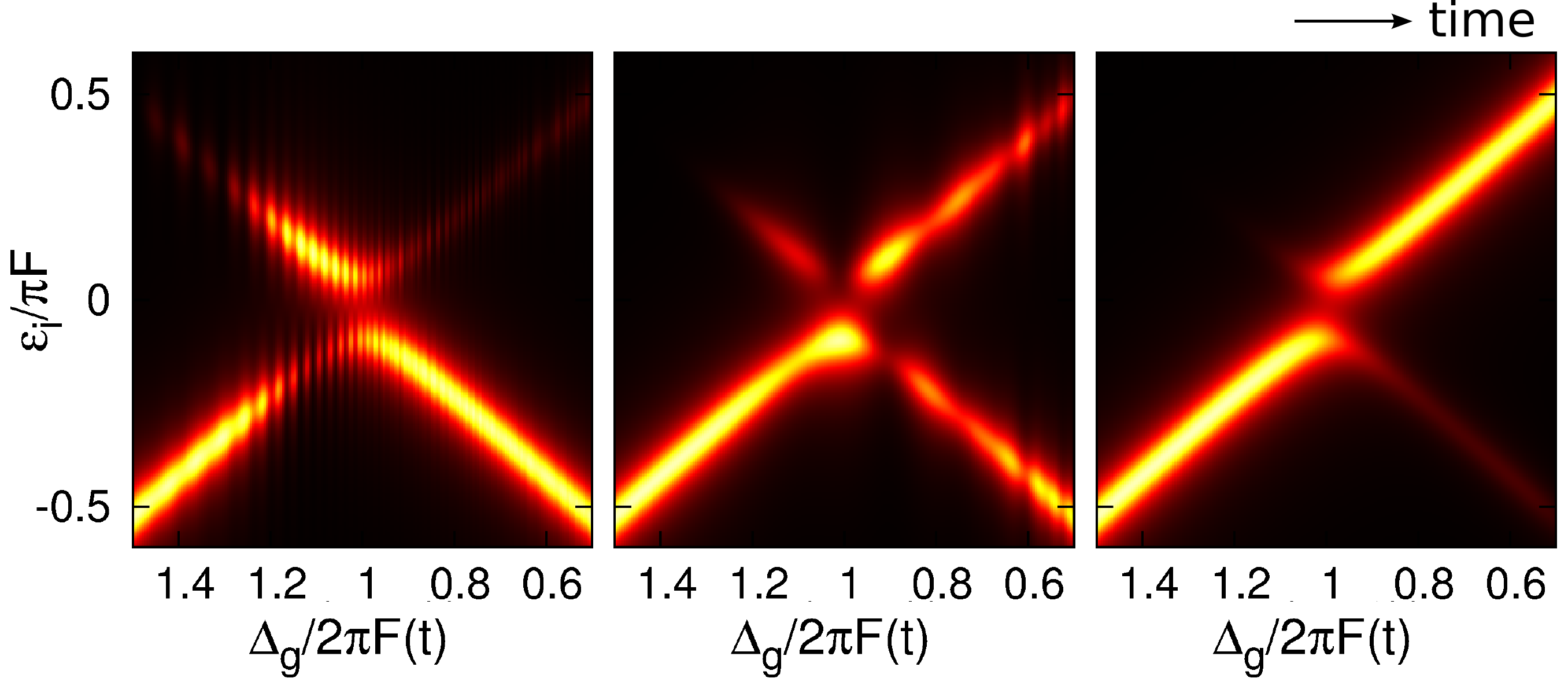}
\caption{\label{fig:3}(Color online)~Driving over the single particle spectrum in Fig.~\ref{fig:2}(a): Parametric-time evolution
(from left to right) of a state initially prepared in the manifold  $M=0$ (bottom left of the panels)
across the resonance $r=1$, with the linear pulse $F(t)=F_0 + \alpha t$. Dynamical parameters: (left) adiabatic transit ($\lambda = 0.1$), (center) non-adiabatic transit ($\lambda=1$) and (right)
diabatic transit ($\lambda = 50$). The color scale is from black (small) to white (large), and
$\delta=0.05$ in Eq.~(\ref{eq:16}). The parameters are the same as in the previous
figure.}
\end{figure}

In figure \ref{fig:3} we show the evolution of the LDOS $P_{\psi}(\varepsilon,t)$ in the single particle case $N=1$, i.e. the 
Hamiltonian $\hat H'_r$ in Eq.~\ref{eq:10} is a $2\times2$ matrix. The initial condition is 
$|\psi(0)\rangle\sim|100,...\rangle_a\otimes|000,...\rangle_b$, $M=0$. $|\psi(0)\rangle$ is evolved from $F_0<F_{r=1}$ to
$F_f>F_{r=1}$ (left to right in Fig.~\ref{fig:2}(a)) for the two-state system $N/L=1/10$,
and we can easily appreciate the different types of dynamics, that is: the left panel shows the adiabatic regime, for
which $|\psi(0)\rangle$ nearly follows the energy path (see low-energy state in Fig.~\ref{fig:2}(a)) while being transformed into 
the excited state with $M=1$. This result is expected according to the exchange of character
typical of an avoided crossing. This is a dynamical effect following from the adiabatic theorem, which was already experimentally probed in
Arimondo's group at Pisa University \cite{WimbPRL2009,Arimondo2011,NilsPRA,Tayebirad2011}. On the other hand, the
central panel shows the non-adiabatic transit for which the probability is split in both energy path. This effect is similar to the
action of a beam splitter on an incident light beam. Finally the right panel depicts the diabatic passage for which the state does preserve its
manifold number $M$ but not its energy path. This latter is characterized by the fidelity of the initial state, i.e. 
$|\langle\psi(0)|\hat U(t)|\psi(0)\rangle|^2\approx 1$.

%******** FIG 2
%******** FIG 3

\section{Numerics: Interaction-induced quantum chaos}\label{sec:4}
%******** FIG 2
\begin{figure}[b]
\centering \includegraphics[width=\columnwidth]{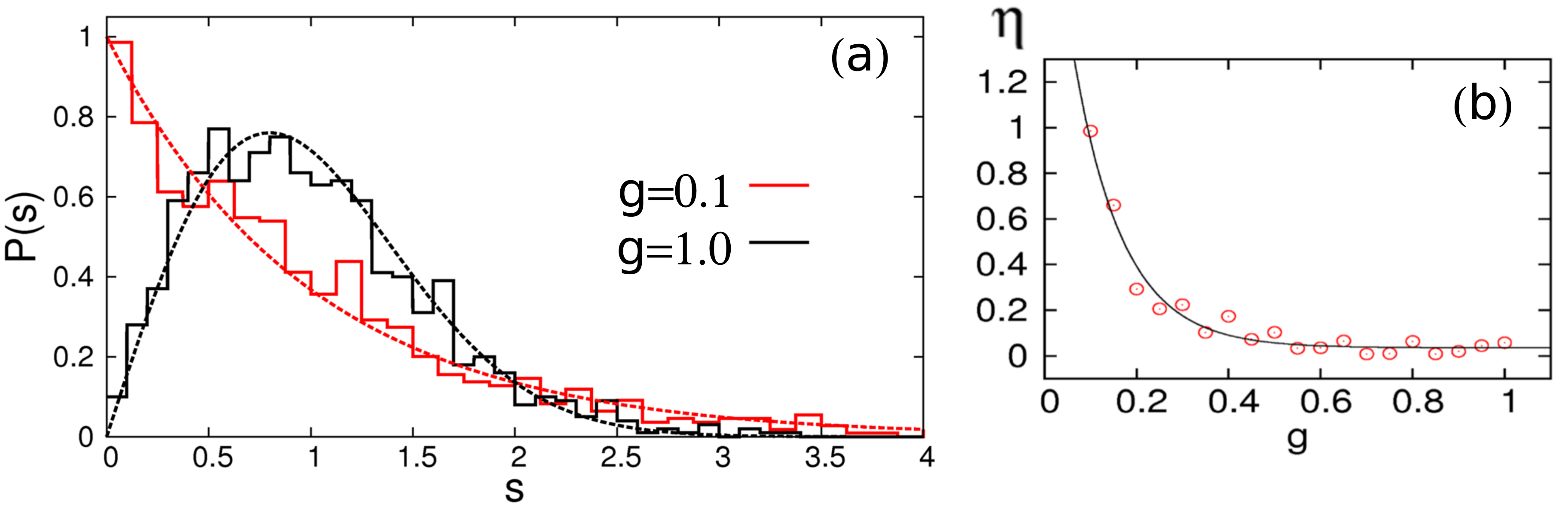}
\caption{\label{fig:4}(Color online)~Interaction effects: (a) The panel shows the level spacing distribution $P(s)$ for $N/L=6/5$ ($\mathcal{N}_s=1001$)
for the interparticle control parameter $g=0.1$ (red/grey histogram, $\eta=0.97$) and $g=1.0$ (black histogram, $\eta=0.035$), and $W_x=W_a=W_b=0.025$.
(b) Parameter $\eta$ as a function of $g$. The black line corresponds to an exponential fit. The reminding parameters are those of Fig.~\ref{fig:2}.}
\end{figure}
%******** FIG 3

In this final section we study the static eigenspectrum at strong mixing conditions. Given the eigenenergies
$\varepsilon_i$, the spectra can be analyzed by the use of random
matrix measures \cite{Haakebook}. A robust test of the universal properties of the discrete eigenspectrum is the so-called level spacing distribution $P(s_i)$, where the 
spacing are defined as $s_i=(\varepsilon_{i+1}-\varepsilon_i)/\langle s\rangle$. The spectrum must be unfolded in order to compare with
the random matrix distributions, i.e., such that $\langle s_i\rangle=1$ (see ref. \cite{CarlosThesis2013} for details). 

At the resonances $F_r$, the crossover between regular (Poisson), $P_P(s)=\exp(-s)$, and quantum chaotic 
(GOE) statistics, $P_W(s)=\pi s\exp(-\pi s^2/4)/2$, is reached by varying the inter-particle interaction. We define the parameter $g$, 
that allows one to tune the interaction, i.e., $W_{\beta,x}\rightarrow gW_{\beta,x}$. In the experiment this can be done via Feschbach resonances \cite{IblochREP2008}.
As expected from the discussion in subsection \ref{sec:2d}, for an energy band gap $\Delta_g\lesssim 1$ and $g=1$, all systems with $N/L\sim 1$
exhibit chaotic features characterized by the GOE distribution (see Fig.~\ref{fig:4}(a)-black). Nevertheless, this is not a general rule for every interaction
strength, since for weakly interacting particles the manifold mixing
becomes weaker. Hence quantum chaos can be tuned most easily by increasing the inter-particle 
interaction, for instance, $g=0\rightarrow 1$. To see this transition we compute the parameter 
\begin{equation}\label{eq:22}
  \eta =\frac{\int^{s_0}_0\left(P(s)-P_W(s)\right)ds}{\int^{s_0}_0\left(P_P(s)-P_W(s)\right)ds},
\end{equation}

\noindent where $s_0=0.4729...$ is the intersection point between $P_P(s)$ and $P_W(s)$. We show the behavior of $\eta$ ($\eta=1$ for a poissonian case
and $\eta=0$ for GOE) as a function of $g$ in fig.~\ref{fig:4}(b).  We see that our two-band model allows us to concentrate a high density of many-body 
energy levels around a resonance, which induces the chaotic features
of the system. This is not the case far from the resonant regime, since here the system is just 
weakly mixed due to the presence of the $M$-manifolds. Then the spectrum is nearly regular \cite{CarlosPRA2013,CarlosThesis2013}, and well characterized by the quantum numbers $M$ and $\vec\theta$ (see Sec.~\ref{sec:2d}), see Fig.~\ref{fig:2}(b).

We thus see that the mixing properties of our system allow for a straightforward identification of the onset of quantum chaos, which is a local effect
in the plane $\varepsilon_i$ vs $F$. Nevertheless, a globally chaotic spectrum can also be reached by decreasing the energy bandgap $\Delta_g$ such 
that $gW_x/\Delta_g\sim 1$. In this case, not only resonant tunneling between two neighboring wells due to the Stark force is possible, but also 
via interaction \cite{MGreiner2011-0,InsbruckArxiv2013}. These
two processes are equally relevant and, as immediate signature of
this, the local resonances are destroyed. There are then avoided
crossings in the entire energy spectrum. Therefore, we
can switch from global to local quantum chaos by varying the bandgap $\Delta_g$.

\section{Conclusions}\label{sec:5}
We studied the spectral properties of a many-body Wannier-Stark system, especially in the resonant tunneling regime. We showed that 
the main characteristics of our model are well described in terms of upper-band excitation subspaces, i.e., $M$-manifolds. This effective
description provides an intuitive understanding of interaction-induced quantum chaos. Furthermore, we have shown that the knowledge
of the spectral properties of the system allows us to control the dynamics when driving a simple two-state system across an avoided
crossing, in the Landau-Zener framework \cite{LZ,Arimondo2011}. Our system can be experimentally realized and it offers an interesting framework
for future investigations of many-body physics \cite{PloetzJPB2010,PloetzEPJD2011,CarlosPRA2013,Oliver-1,InsbruckArxiv2013}.

\section{Acknowledgments}

We acknowledge financial support from the DFG (FOR760), the HGSFP
(GSC 129/1) and the Institute for Theoretical Physics at Heidelberg
University. It is our pleasure to warmly thank J. Madro\~nero for
lively discussions. S.W. thanks furthermore the organizers of the workshop,
L. Sirko and S. Bauch, and M. Ku\'s for his kind hospitality at the
Polnish Academy of Sciences in Warsaw.

%\newpage

\end{document}